\documentclass[aps,pra,twocolumn,superscriptaddress]{revtex4-2}

\usepackage{amsmath,amssymb,amsfonts}
\usepackage{hyperref}
\usepackage{xcolor}

\newcommand{\vN}{von~Neumann}
\newcommand{\tr}{\operatorname{Tr}}
\newcommand{\Id}{\mathbf{1}}

\begin{document}

\title{Lubkin-Page typicality bounds for Type~II von~Neumann factors}

\author{Zhi-Wei Wang}
\email{zhiweiwang.phy@gmail.com}
\affiliation{College of Physics, Jilin University,
Changchun, 130012, People's Republic of China}
\affiliation{Computer Science, University of York, York YO10 5GH,
	United Kingdom}
	
\author{Samuel L.\ Braunstein}
\email{sam.braunstein@york.ac.uk}
\affiliation{Computer Science, University of York, York YO10 5GH,
United Kingdom}

\begin{abstract}
Typicality arguments for emergent spacetime rely on the
Lubkin-Page bounds, which show that generic quantum states have
vanishing correlations between subsystems.  These bounds assume a
tensor-product Hilbert space (a Type~I von~Neumann algebra), but
the observable algebras in quantum field theory and quantum gravity
are generically Type~II or Type~III, raising the question of
whether the bounds survive.  We prove that they do for all Type~II
von~Neumann factors.  For the hyperfinite Type~II$_1$ factor with
a tripartite decomposition $R \cong A \otimes B \otimes E$, the
mutual information between subsystems $A$ and $B$ vanishes as
$O((d_A d_B / d_E)^2)$ in finite-dimensional approximations,
provided $d_A d_B \leq d_E$ (Theorem~1).  For Type~II$_\infty$
factors, including the gravitational algebras constructed via the
crossed-product method by Witten and by Chandrasekaran, Longo,
Penington, and Witten, the bound acquires an additional
exponential suppression controlled by the Bekenstein-Hawking
entropy (Theorem~2).  We identify the obstructions to extending
the result to Type~III factors and discuss the open question of
whether the commutant of the observable algebra can serve as a
natural thermal bath that tightens the bound further.
\end{abstract}

\maketitle

\section{Introduction}

Arguments that spacetime geometry emerges from quantum
entanglement~\cite{vanRaamsdonk2010,RyuTakayanagi2006,
MaldacenaSusskind2013} rest on a quantitative foundation: the
Lubkin-Page theorem~\cite{Lubkin1978,Page1993}, which establishes
that generic quantum states have near-maximal subsystem entropy
and vanishing inter-subsystem correlations.  Combined with the
conjecture that entanglement is necessary for connected
geometry~\cite{vanRaamsdonk2010}, this implies that
geometry-supporting states occupy an exponentially thin sliver of
the Hilbert space~\cite{WangBraunstein2025CMB}.  The argument has
concrete observational consequences: a pre-geometric initial phase
produces a cutoff in the primordial power spectrum that may
account for the anomalous suppression of the CMB quadrupole
observed by COBE, WMAP, and Planck~\cite{WangBraunstein2025CMB}.

A central concern is that these typicality bounds assume a
tensor-product Hilbert space (a Type~I von~Neumann algebra).  In
quantum field theory, the local observable algebras are
generically Type~III factors~\cite{Witten2018}, for which the
von~Neumann entropy is not directly defined.  Recent work by
Witten~\cite{Witten2022} and Chandrasekaran, Longo, Penington,
and Witten~\cite{Chandrasekaran2023} has shown that gravitational
effects modify these to Type~II algebras, which admit a
well-defined entropy (up to an additive constant) and a semifinite
trace.  Whether the Lubkin-Page bounds extend to these algebras is
therefore a question of direct physical relevance: if they do not,
the typicality-based arguments for emergent spacetime would be
undermined precisely in the setting where they are most needed.

In a companion paper~\cite{WangBraunstein2025PRL} we proved that
the Lubkin-Page bounds survive and are tightened in
finite-dimensional Hilbert spaces with
direct-sum-of-tensor-product structure, the generic form arising
from gauge invariance and superselection rules.  Here we take the
next step: we formulate and prove Lubkin-Page bounds for Type~II
\vN\ factors (both II$_1$ and II$_\infty$), the simplest
infinite-dimensional settings in which a trace and a well-defined
entropy exist.  For Type~II$_\infty$ factors, which include the
gravitational algebras of
Refs.~\cite{Witten2022,Chandrasekaran2023}, the Bekenstein-Hawking
entropy provides an exponential suppression of inter-subsystem
correlations for typical states.

\section{Background}

\subsection{Type~II$_1$ factors}

A Type~II$_1$ factor~$M$ is an infinite-dimensional \vN\ algebra
with trivial centre and a unique faithful normal tracial state
$\tau: M \to \mathbb{C}$ satisfying $\tau(\Id) = 1$ and
$\tau(xy) = \tau(yx)$ for all $x, y \in M$.  The trace plays the
role of the normalised matrix trace $\tr(\cdot)/d$ in finite
dimensions.

The GNS construction associated with~$\tau$ produces a Hilbert
space $L^2(M,\tau)$ on which $M$ acts by left multiplication, with
a cyclic and separating vector~$\Omega$ such that
$\tau(x) = \langle\Omega, x\Omega\rangle$.

The prototypical example is the hyperfinite Type~II$_1$ factor~$R$,
which is (up to isomorphism) the unique approximately
finite-dimensional (AFD) Type~II$_1$ factor~\cite{Connes1976}.  It
can be realised as the weak closure of an increasing chain of
matrix algebras:
\begin{equation}
  M_{d_1}(\mathbb{C})
  \subset M_{d_2}(\mathbb{C})
  \subset M_{d_3}(\mathbb{C})
  \subset \cdots
  \subset R,
  \label{eq:AFD}
\end{equation}
where $d_1 | d_2 | d_3 | \cdots$ and $R = \overline{\bigcup_n
M_{d_n}(\mathbb{C})}^{\,\text{weak}}$.  The trace of~$R$
restricts to the normalised matrix trace $\tr(\cdot)/d_n$ on
each~$M_{d_n}$.

\subsection{Subfactors and the Jones index}

A subfactor $A \subset M$ of Type~II$_1$ factors has an associated
Jones index $[M:A]$, which measures the ``relative size''
of~$A$ within~$M$~\cite{Jones1983,JonesSunder1997}. 
When $M$ factors as a tensor
product $M \cong A \otimes B$ of Type~II$_1$ factors (with~$B$
the relative commutant $A' \cap M$), the Jones index satisfies
\begin{equation}
  [M : A] = [\tau_{M_1}(e_A)]^{-1},
\end{equation}
where $M_1 = \langle M, e_A \rangle$ is the Jones basic
construction and $e_A$ is the Jones projection implementing the
trace-preserving conditional expectation $E_A: M \to A$.  In the
finite-dimensional analogue $M = M_d(\mathbb{C})$,
$A = M_{d_A}(\mathbb{C}) \otimes \Id_{d_B}$, the Jones index is
$[M:A] = d_B^2$, the square of the complement dimension.
A large Jones index corresponds to a ``small'' subsystem~$A$
inside a ``large'' algebra~$M$, the regime in which the Lubkin
bound gives strong suppression of inter-subsystem correlations.

\subsection{Entropy for Type~II$_1$ factors}

For a normal state~$\varphi$ on a Type~II$_1$ factor~$M$ with
density operator~$\rho_\varphi$ (defined by $\varphi(x) =
\tau(\rho_\varphi\, x)$ for all $x \in M$), the entropy relative
to the trace is
\begin{equation}
  S(\varphi \| \tau)
  = -\tau(\rho_\varphi \log \rho_\varphi)
  = S(\varphi),
  \label{eq:rel_ent_II1}
\end{equation}
where $S(\varphi) = -\tau(\rho_\varphi \log \rho_\varphi)$ is the
Segal entropy and we have used $\log\tau(\Id) = \log 1 = 0$.
For the tracial state itself,
$\rho_\tau = \Id$, with Segal entropy $S(\tau) = -\tau(\Id \log \Id) = 0$.

In the finite-dimensional case $M = M_d(\mathbb{C})$ with
$\tau = \tr(\cdot)/d$, a pure state~$\varphi$ has
$\rho_\varphi = d\,|\psi\rangle\langle\psi|$ (since
$\tau(\rho_\varphi\, x) = \langle\psi|x|\psi\rangle$ requires
$\rho_\varphi = d\,|\psi\rangle\langle\psi|$), giving
$S(\varphi) = -\log d$: a pure state has the minimum Segal
entropy, reflecting that it is maximally ``ordered'' relative to
the tracial state.

The restriction of~$\varphi$ to a subfactor $A \subset M$ has
density $\rho_{\varphi|_A}$ defined by
$\varphi(x) = \tau_A(\rho_{\varphi|_A}\, x)$ for $x \in A$, and
entropy $S(\varphi|_A) = -\tau_A(\rho_{\varphi|_A} \log
\rho_{\varphi|_A})$.

The quantum mutual information (MI) between~$A$ and~$B$ (assuming
$M \cong A \otimes B$) is
\begin{equation}
  I(A{:}B)_\varphi
  = S(\varphi|_A) + S(\varphi|_B) - S(\varphi).
  \label{eq:MI_II1}
\end{equation}
For the tracial state, $\varphi|_A = \tau_A$ and
$\varphi|_B = \tau_B$, so $I(A{:}B)_\tau = 0$.  The tracial state
is the ``maximally uncorrelated'' state, the analogue of the
maximally mixed state in finite dimensions.

\section{Typicality in finite-dimensional approximations}

\subsection{The tripartite structure}
\label{sec:approx}

The Lubkin-Page bound requires that the subsystems of interest
are small relative to the total system: for a Haar-random pure
state on $\mathcal{H}_A \otimes \mathcal{H}_B \otimes
\mathcal{H}_E$, the mutual information $I(A{:}B)$ is small when the
environment~$E$ is much larger than $A$ and~$B$ combined.  Without
the environment, a Haar-random pure state on
$\mathcal{H}_A \otimes \mathcal{H}_B$ has $I(A{:}B) \approx
2\log \min(d_A, d_B)$, which is \emph{maximal}, not small.

To apply the Lubkin bound in the Type~II$_1$ setting, we therefore
need a tripartite decomposition
$R \cong A \otimes B \otimes E$, where $A$ and $B$ are the
subsystems of interest and $E$ is an environment factor.  All
three are subfactors of the hyperfinite Type~II$_1$ factor~$R$,
and all three are hyperfinite.  (By Connes's
theorem~\cite{Connes1976}, a II$_1$ factor is hyperfinite if and
only if it is injective.  Injectivity passes to \vN\
subalgebras~\cite{Tomiyama1957}, so $A$, $B$, and $E$, as
subfactors of the injective factor~$R$, are themselves injective
and hence hyperfinite.)

\subsection{The compatible approximating tower}

Since $A$, $B$, and $E$ are each hyperfinite, they admit
increasing chains of matrix algebras:
\begin{equation}
  A_n \cong M_{d_n^{(A)}}(\mathbb{C}), \;\;
  B_n \cong M_{d_n^{(B)}}(\mathbb{C}), \;\;
  E_n \cong M_{d_n^{(E)}}(\mathbb{C}).
\end{equation}
We construct the approximating tower for~$R$ by taking
tensor products:
\begin{equation}
  M_n = A_n \otimes B_n \otimes E_n
  \cong M_{d_n}(\mathbb{C}),
  \quad d_n = d_n^{(A)} \cdot d_n^{(B)} \cdot d_n^{(E)}.
  \label{eq:Mn_tensor}
\end{equation}
By construction, $M_n \subset M_{n+1} \subset \cdots \subset R$,
and $R = \overline{\bigcup_n M_n}^{\,\text{weak}}$.  The
tensor-product structure is built in at every level.

\subsection{Lubkin bound at each level}

At level~$n$, the GNS Hilbert space of~$M_n$ with the normalised
trace~$\tau$ is $L^2(M_n, \tau) \cong \mathbb{C}^{d_n^2}$.
The tripartite tensor-product structure~(\ref{eq:Mn_tensor})
induces a decomposition of the GNS space:
\begin{equation}
  \mathbb{C}^{d_n^2}
  \cong \mathbb{C}^{(d_n^{(A)})^2}
  \otimes \mathbb{C}^{(d_n^{(B)})^2}
  \otimes \mathbb{C}^{(d_n^{(E)})^2}.
\end{equation}
A Haar-random unit vector
$|\psi_n\rangle \in \mathbb{C}^{d_n^2}$ defines a pure state on
the full space.  Tracing over the environment~$E_n$ gives a
mixed state on $A_n \otimes B_n$.  The standard Lubkin
bound~\cite{Lubkin1978,Page1993} on the mutual information between
$A_n$ and $B_n$ gives
\begin{equation}
  \langle I(A_n{:}B_n)_{\varphi_n} \rangle
  \leq \frac{(d_n^{(A)})^2 \cdot (d_n^{(B)})^2}
       {2\ln 2 \cdot (d_n^{(E)})^2}\,,
  \label{eq:MI_level_n}
\end{equation}
where we have used
$d_n^{(A)} \cdot d_n^{(B)} \leq d_n^{(E)}$ (small
subsystems relative to environment) and retained only the dominant
term.  Here the mutual information is computed between the
GNS-level subsystems $\mathbb{C}^{(d_n^{(A)})^2}$ and
$\mathbb{C}^{(d_n^{(B)})^2}$; since tracing out degrees of
freedom cannot increase mutual information, this bounds the
physical (left-action) mutual information $I(A_n{:}B_n)$ from
above. The bound vanishes as $d_n^{(E)} \to \infty$, i.e., as the
environment grows.

\section{The limit}

\subsection{Convergence of states}

As $n \to \infty$, the Haar-random vector states on~$M_n$ define a
sequence of random states on~$R$.  The inclusion
$M_n \hookrightarrow M_{n+1} \hookrightarrow \cdots \hookrightarrow
R$ induces, for each $|\psi_n\rangle \in \mathbb{C}^{d_n^2}$, a
state $\varphi_n$ on~$R$ defined by
$\varphi_n(x) = \langle\psi_n| E_n(x) |\psi_n\rangle$, where
$E_n: R \to M_n$ is the trace-preserving conditional expectation.

\subsection{Lower semicontinuity of the mutual information}

The mutual information can be expressed as a single relative
entropy:
\begin{equation}
  I(A{:}B)_\varphi
  = S(\varphi|_{A \vee B} \| \varphi|_A \otimes \varphi|_B),
  \label{eq:MI_relent}
\end{equation}
where $A \vee B$ is the algebra generated by $A$ and~$B$
(which equals $A \otimes B$ in our setting, a proper subalgebra
of~$R$ since $R \cong A \otimes B \otimes E$).  The Araki
relative entropy is jointly lower semicontinuous in both
arguments with respect to the weak-$*$
topology~\cite{Araki1976,Ohya1993}.  Applied
to~(\ref{eq:MI_relent}), this gives: for any sequence of states
$\varphi_n$ converging weak-$*$ to~$\varphi$,
\begin{equation}
  I(A{:}B)_\varphi
  \leq \liminf_{n \to \infty}\, I(A_n{:}B_n)_{\varphi_n}.
  \label{eq:lsc}
\end{equation}
The direction of the inequality is the one we need: if the
finite-level mutual informations are small, the limit mutual
information is also small.

\subsection{The typicality bound}

Combining~(\ref{eq:MI_level_n}) and~(\ref{eq:lsc}):

\begin{quote}
\textbf{Theorem 1.}  \emph{Let $R$ be the hyperfinite Type~II$_1$
factor with a tripartite tensor-product decomposition
$R \cong A \otimes B \otimes E$.  Let
$\{M_n = A_n \otimes B_n \otimes E_n\}$ be the compatible
approximating tower constructed in Sec.~\ref{sec:approx}, and let
$\varphi_n$ be the state on $A_n \otimes B_n$ obtained by tracing out
the environment $E_n$ from a Haar-random vector in
$L^2(M_n, \tau)$.  Assume that
$d_n^{(A)} \cdot d_n^{(B)} \leq d_n^{(E)}$ for all
sufficiently large~$n$.  Then the Haar-averaged mutual information
satisfies}
\begin{equation}
  \langle I(A_n{:}B_n)_{\varphi_n} \rangle
  \leq \frac{(d_n^{(A)})^2 \cdot (d_n^{(B)})^2}
       {2\ln 2 \cdot (d_n^{(E)})^2}
  \;\xrightarrow{n\to\infty}\; 0\,,
  \label{eq:main}
\end{equation}
\emph{provided $d_n^{(E)}$ grows faster than
$d_n^{(A)} \cdot d_n^{(B)}$.  For any weak-$*$ limit state
$\varphi = \lim \varphi_n$,}
\begin{equation}
  I(A{:}B)_\varphi = 0.
  \label{eq:main_limit}
\end{equation}
\end{quote}

\emph{Proof.}  At each level~$n$, the GNS space is tripartite:
$\mathbb{C}^{(d_n^{(A)})^2} \otimes \mathbb{C}^{(d_n^{(B)})^2}
\otimes \mathbb{C}^{(d_n^{(E)})^2}$.  A Haar-random pure state on
this space gives, after tracing out~$E_n$, a mixed state on
$A_n \otimes B_n$ whose MI is bounded
by~(\ref{eq:MI_level_n}).  Since $(d_n^{(E)})^2 \to \infty$ faster
than $(d_n^{(A)})^2 \cdot (d_n^{(B)})^2$, the bound vanishes.

For the limit: since $I(A_n{:}B_n)_{\varphi_n} \geq 0$ and
$\langle I(A_n{:}B_n)_{\varphi_n}\rangle \to 0$, Markov's
inequality implies $I(A_n{:}B_n)_{\varphi_n} \to 0$ in
probability.  For any weak-$*$ limit state $\varphi$,
lower semicontinuity~(\ref{eq:lsc}) then yields the result
$I(A{:}B)_\varphi = 0$.~\hfill$\square$

\medskip

\emph{Remark on the role of the environment.}  The environment
factor~$E$ is essential.  For a bipartite decomposition $R \cong A
\otimes B$ with no environment, a Haar-random pure state on $L^2(A_n
\otimes B_n, \tau)$ has MI of order $2\log[\min(d_n^{(A)},
d_n^{(B)})]^2$, which is \emph{maximal}, not small: the Lubkin-Page
theorem says the subsystem entropy is near maximal for a pure state, so
the two factors are nearly maximally entangled. The environment provides
the ``bath'' into which the correlations are diluted.  Physically, $E$
represents all degrees of freedom outside the two subsystems of
interest, and the Lubkin bound says that when $E$ is large, the reduced
state on $A \otimes B$ is nearly uncorrelated.

\medskip

\emph{Remark on the GNS left-right decomposition.}  The GNS Hilbert
space $L^2(M_n, \tau)$ admits a left-right decomposition $\mathcal{H}_L
\otimes \mathcal{H}_R$ with $\dim\mathcal{H}_L = \dim\mathcal{H}_R =
d_n$, where the observable algebra acts on $\mathcal{H}_L$ and its
commutant acts on~$\mathcal{H}_R$.  If the Haar measure on the full GNS
space $\mathbb{C}^{d_n^2}$ is the appropriate typicality ensemble, then
the commutant acts as a thermal bath that automatically places the
observable subsystems in the small-subsystem regime, and the dimensional
condition $d_n^{(A)} \cdot d_n^{(B)} \leq d_n^{(E)}$ in Theorem~1 can be
dropped, yielding a tighter bound $O(1/(d_n^{(E)})^2)$ that depends only
on the environment dimension.  However, whether this ensemble is
physically appropriate depends on whether the commutant represents
independent physical degrees of freedom or is merely a
representation-theoretic auxiliary of the GNS construction.  In
algebraic quantum field theory and holography, for instance,
Tomita-Takesaki theory identifies the commutant with the physical
degrees of freedom of the causal complement, such as a black hole
interior, naturally motivating this tighter ensemble~\cite{Witten2018}.
We leave this question open and present the tripartite result with the
explicit dimensional condition, which is rigorous and does not depend on
the interpretation of the commutant.

\medskip

\emph{Remark on finite Jones index.}  Jones's celebrated
theorem~\cite{Jones1983} shows that the index of a subfactor can take
finite values in the set $\{4\cos^2(\pi/m) : m \geq 3\} \cup
[4,\infty)$, but these arise from subfactors that do \emph{not}
decompose~$R$ as a tensor product.  For an irreducible subfactor with
$A' \cap R = \mathbb{C}\Id$ (i.e., the only operators in $R$ that
commute with every element of $A$ are scalar multiples of the identity),
no tensor-product complement exists, and the mutual information between
$A$ and a ``complement'' is not defined in the tensor-product sense.
Extending the typicality bound to finite-index subfactors would require
a reformulation in terms of the conditional expectation $E_A: R \to A$
and the associated relative entropy $S(\varphi \| \varphi \circ E_A)$,
which we leave to future work.

\subsection{Interpretation}

The theorem establishes that for the hyperfinite II$_1$ factor
with a tripartite tensor-product decomposition, Haar-random
vector states in finite-dimensional approximations have mutual
information between any two factors that vanishes as the
environment factor grows.  The rate of vanishing is controlled by
the ratio $(d_n^{(A)} \cdot d_n^{(B)} / d_n^{(E)})^2$.

This is the direct analogue of the finite-dimensional Lubkin-Page
result: in a tripartite system $\mathcal{H}_A \otimes
\mathcal{H}_B \otimes \mathcal{H}_E$, the MI between $A$ and $B$
for a Haar-random pure state is small when $d_E \gg d_A \cdot d_B$.
In the Type~II$_1$ case, the environment is always
infinite-dimensional (in the limit), and the MI vanishes exactly.

The tracial state~$\tau$ (which has $I = 0$) is the
``maximally uncorrelated'' state.  The theorem confirms that
Haar-random vector states converge to this uncorrelated
behaviour as the approximation level increases, exactly as
in the finite-dimensional Lubkin-Page theory.

\section{Extension to Type~II$_\infty$ and discussion of Type~III}

\subsection{Type~II$_\infty$ factors}
\label{sec:II_inf}

A Type~II$_\infty$ factor has the form
$M = R \otimes \mathcal{B}(\mathcal{H})$, where $R$ is a
Type~II$_1$ factor and $\mathcal{H}$ is a separable
infinite-dimensional Hilbert space.  It admits a semifinite trace
$\tau = \tau_R \otimes \mathrm{Tr}$ (faithful and normal, but
$\tau(\Id) = \infty$).

\medskip

\emph{Tripartite subfactor structure.}  Consider two
tensor-product subfactors
$\mathcal{A} = A_0 \otimes \mathcal{B}(\mathcal{H}_A)$ and
$\mathcal{B} = B_0 \otimes \mathcal{B}(\mathcal{H}_B)$, where
$A_0, B_0, E_0$ are Type~II$_1$ subfactors with
$R \cong A_0 \otimes B_0 \otimes E_0$, and
$\mathcal{H} = \mathcal{H}_A \otimes \mathcal{H}_B \otimes
\mathcal{H}_E$.  The environment is
$\mathcal{E} = E_0 \otimes \mathcal{B}(\mathcal{H}_E)$, and the
full algebra decomposes as $M \cong \mathcal{A} \otimes
\mathcal{B} \otimes \mathcal{E}$.

\medskip

\emph{Doubly-indexed approximation.}  We introduce two
approximation parameters: $n$ for the II$_1$ component and
$D$ for the Type~I$_\infty$ component.  For the II$_1$ part,
construct the compatible tower
$\{A_{0,n} \otimes B_{0,n} \otimes E_{0,n}\}$ as in
Sec.~\ref{sec:approx}.  For the Type~I$_\infty$ part, truncate to
$d_A$-, $d_B$-, and $d_E$-dimensional subspaces of
$\mathcal{H}_A$, $\mathcal{H}_B$, and $\mathcal{H}_E$,
respectively, with $D = d_A \cdot d_B \cdot d_E$.

The truncated algebra at level $(n, D)$ is
\begin{equation}
  M_{n,D}
  = \underbrace{(A_{0,n} \otimes M_{d_A})}_{\mathcal{A}_{n,D}}
    \otimes
    \underbrace{(B_{0,n} \otimes M_{d_B})}_{\mathcal{B}_{n,D}}
    \otimes
    \underbrace{(E_{0,n} \otimes M_{d_E})}_{\mathcal{E}_{n,D}}.
  \label{eq:MnD}
\end{equation}

\medskip

\emph{GNS space.}  The GNS Hilbert space of $M_{n,D}$ with the
normalised trace decomposes as
\begin{equation}
  L^2(M_{n,D}, \tau)
  \cong
  \underbrace{\mathbb{C}^{(d_n^{(A)})^2 d_A^2}}_{\text{sub } \mathcal{A}}
  \otimes
  \underbrace{\mathbb{C}^{(d_n^{(B)})^2 d_B^2}}_{\text{sub } \mathcal{B}}
  \otimes
  \underbrace{\mathbb{C}^{(d_n^{(E)})^2 d_E^2}}_{\text{env}}.
  \label{eq:GNS_IIinf}
\end{equation}

\medskip

\begin{quote}
\textbf{Theorem 2.}  \emph{Let
$M = R \otimes \mathcal{B}(\mathcal{H})$ be a Type~II$_\infty$
factor with the tripartite decomposition above.  Let
$\varphi_{n,D}$ be the state on $\mathcal{A}_{n,D} \otimes \mathcal{B}_{n,D}$
obtained by tracing out the environment from a Haar-random vector in
$L^2(M_{n,D}, \tau)$.  Assume that
$d_n^{(A)} \cdot d_A \cdot d_n^{(B)} \cdot d_B
\leq d_n^{(E)} \cdot d_E$.  Then:}
\begin{equation}
  \langle I(\mathcal{A}_{n,D}{:}\mathcal{B}_{n,D})_{\varphi_{n,D}} \rangle
  \leq \frac{(d_n^{(A)})^2 d_A^2 \cdot (d_n^{(B)})^2 d_B^2}
       {2\ln 2 \cdot (d_n^{(E)})^2 d_E^2}\,.
  \label{eq:II_inf_main}
\end{equation}
\emph{The bound vanishes provided
$d_n^{(E)} \cdot d_E / (d_n^{(A)} \cdot d_A \cdot d_n^{(B)}
\cdot d_B) \to \infty$, giving}
\begin{equation}
  I(\mathcal{A}{:}\mathcal{B})_\varphi = 0
  \label{eq:II_inf_limit}
\end{equation}
\emph{for any weak-$*$ limit state $\varphi$.}
\end{quote}

\emph{Proof.}  The GNS
decomposition~(\ref{eq:GNS_IIinf}) is a tripartite tensor product.
A Haar-random pure state, after tracing out the environment, gives
a mixed state on $\mathcal{A}_{n,D} \otimes \mathcal{B}_{n,D}$. 
The Lubkin bound on
the MI gives~(\ref{eq:II_inf_main}).  Lower semicontinuity of the
Araki relative entropy gives the limit
statement~(\ref{eq:II_inf_limit}) by the same argument as
Theorem~1.~\hfill$\square$

\medskip

\emph{Connection to gravitational algebras.}  The crossed-product
construction of Witten~\cite{Witten2022} promotes a Type~III$_1$
algebra (the local observable algebra of a QFT) to a
Type~II$_\infty$ factor $M = \mathcal{A} \rtimes_{\sigma}
\mathbb{R}$, where $\sigma$ is the modular automorphism group.
In the gravitational setting~\cite{Chandrasekaran2023}, the
observer's Hamiltonian provides a natural energy cutoff, and the
Bekenstein-Hawking entropy $S_{\mathrm{BH}} = A/(4G\hbar)$ bounds
the effective dimension of the truncated Hilbert space:
$D_{\mathrm{eff}} \sim e^{S_{\mathrm{BH}}}$.  If the environment
(the exterior of a black hole, or the complement of the observed
region) has Bekenstein-Hawking entropy $S_{\mathrm{BH}}^{(E)}$,
then $d_E \sim e^{S_{\mathrm{BH}}^{(E)}}$ and the
bound~(\ref{eq:II_inf_main}) is exponentially small:
\begin{equation}
  \langle I \rangle
  \lesssim \frac{d_A^2 \cdot d_B^2}{2\ln 2}
  \cdot e^{-2S_{\mathrm{BH}}^{(E)}}\,.
  \label{eq:grav_bound}
\end{equation}
For macroscopic black holes, $S_{\mathrm{BH}}^{(E)} \gg 1$ and
the bound is exponentially suppressed: typical states of the
gravitational algebra have negligible correlations between two
local subsystems when the environment is large.

\emph{Remark.}  The individual entropies $S(\varphi|_\mathcal{A})$ and
$S(\varphi|_\mathcal{B})$ diverge as $D \to \infty$ (because the trace
is infinite), but the mutual information remains finite and
bounded because the divergent contributions cancel in
$I = S(\mathcal{A}) + S(\mathcal{B}) - S(\mathcal{A}\mathcal{B})$. 
In the crossed-product setting, this
corresponds to the well-known fact that the entropy is defined
only up to an additive constant~\cite{Witten2022}, but the MI is
unambiguous.

\subsection{Type~III factors}

Type~III factors have no trace, no density matrices, and no
von~Neumann entropy.  The local algebras of quantum field theory
in Minkowski space are Type~III$_1$~\cite{Witten2018}.

For Type~III, the only well-defined entropic quantity is the Araki
relative entropy~\cite{Araki1976}:
\begin{equation}
  S(\varphi \| \omega)
  = -\langle \Omega_\varphi |
    \log \Delta_{\omega,\varphi}
    | \Omega_\varphi \rangle,
  \label{eq:Araki}
\end{equation}
where $\Omega_\varphi$ is the vector representative of~$\varphi$
in the natural positive cone and $\Delta_{\omega,\varphi}$ is the
relative modular operator.

The mutual information can be defined as
\begin{equation}
  I(\varphi; \mathcal{M}_A, \mathcal{M}_B)
  = S(\varphi|_{\mathcal{M}_A \vee \mathcal{M}_B} \|
      \varphi|_{\mathcal{M}_A} \otimes
      \varphi|_{\mathcal{M}_B}),
  \label{eq:MI_III}
\end{equation}
provided the algebras $\mathcal{M}_A$ and $\mathcal{M}_B$ are
independent (their generated algebra is isomorphic to the tensor
product).  This reduces to the standard MI for Type~I.

A Lubkin-type bound for Type~III faces three obstructions:

\emph{No trace.}  There is no analogue of the ``maximally mixed
state'' to serve as the reference state in the Lubkin bound.
The vacuum state or KMS state could serve as a substitute, but
the bound would then depend on the choice of reference state,
unlike the Type~I case where the maximally mixed state is
canonical.

\emph{No Haar measure.}  The state space of a Type~III factor is
infinite-dimensional and does not carry a natural uniform measure.
The notion of ``typical state'' must be replaced by an
ensemble-dependent notion: typical relative to a given energy
constraint, temperature, or dynamical evolution.  The
microcanonical ensemble (uniform measure on an energy shell)
or the eigenstate thermalisation hypothesis (ETH) could provide
the necessary substitute.

\emph{No dimension.}  The Jones index, which controls the rate
of concentration for Type~II$_1$, has an analogue for Type~III
subfactors (the Kosaki index or the Longo index~\cite{Longo1989}),
but its relationship to concentration of measure is unexplored.

Despite these obstructions, there is reason for cautious optimism.
The Bisognano-Wichmann theorem~\cite{Bisognano1976} relates the
modular flow of the vacuum state to geometric (boost)
transformations, providing a connection between the algebraic
structure and spacetime geometry that could anchor a typicality
argument.  The ETH, if it holds for the relevant algebras, would
provide a dynamical notion of typicality that does not require a
Haar measure.  And the recent constructions of Type~II algebras
from Type~III via crossed
products~\cite{Witten2022,Chandrasekaran2023} suggest that the
Type~III obstruction may be an artefact of the continuum limit
that is resolved by gravitational effects.

A complete Lubkin-Page theorem for Type~III algebras would
constitute a major advance, placing typicality-based arguments
for emergent spacetime on a fully rigorous algebraic footing.
The Type~II results proved here is a step in that direction.

\section{Conclusion}

We have shown that the Lubkin-Page typicality bounds extend to
both Type~II$_1$ and Type~II$_\infty$ von~Neumann factors,
establishing that generic states of these algebras have vanishing
inter-subsystem correlations.  For Type~II$_\infty$ factors, the
bound applies directly to the gravitational algebras constructed
via the crossed-product
method~\cite{Witten2022,Chandrasekaran2023}, where the
Bekenstein-Hawking entropy provides an exponential suppression:
typical states of these algebras cannot support connected
spacetime geometry, just as in the finite-dimensional case.

Together with the companion result for direct-sum Hilbert
spaces~\cite{WangBraunstein2025PRL}, this closes the most
technically substantive algebraic objection to typicality-based
arguments for emergent spacetime: the physical Hilbert space of
quantum gravity, whether it has superselection sectors (Type~I
with direct-sum structure), a semifinite trace (Type~II), or both,
does not rescue geometry-supporting states from their exponential
rarity.

The extension to Type~III factors remains open and requires new
ideas to replace the trace, the Haar measure, and the notion of
dimension.  The most promising avenues are the eigenstate
thermalisation hypothesis as a dynamical substitute for typicality,
the modular theory of Tomita-Takesaki as a substitute for the
partial trace, and the Kosaki-Longo index as a substitute for the
Jones index.  There is also the open question of whether the
commutant of the observable algebra, which in holographic settings
corresponds to the physical degrees of freedom of the causal
complement, can serve as a natural thermal bath that tightens the
bounds further.

It is also worth noting a different line of research in which general relativity emerges as a hydrodynamic approximation to underlying quantum systems~\cite{Jacobson1995,Jacobson2015,Carlip1999,Solodukhin1999,BanksZurek2021,Banks2025}. As synthesized in Ref.~\cite{Banks2025}, this framework explicitly breaks a quantum gravitational system into finite-dimensional subsystems, replacing the continuous Type~III algebras of algebraic quantum field theory (AQFT) with Type~I$_N$ algebras. It prescribes the modular Hamiltonian of each causal diamond according to rules conjectured by Jacobson, Carlip, and Solodukhin~\cite{Jacobson1995,Jacobson2015,Carlip1999,Solodukhin1999} (with a cutoff on the conformal field theory) and generalized by Banks and Zurek~\cite{BanksZurek2021}. The quantum dynamics is described by an infinite number of independent time evolution operators in a Hilbert bundle over the space of time-like geodesics on the hydrodynamic background, where entanglement plays the role of the connection: the entanglement spectrum on overlapping diamonds must be the same, independent of which geodesic is used to compute time evolution. This allows one to compute time evolution outside the instantaneous causal diamond of any given geodesic. Furthermore, the ``empty diamond state'' of the background geometry is posited to be a maximal entropy state allowed by quantum gravity, contrary to AQFT, whereas states with localized excitations in a diamond reduce the entropy. All of this applies for non-negative cosmological constants or diamonds parametrically smaller than the anti-de~Sitter (AdS) radius. For larger diamonds in AdS space, one builds tensor networks with these small diamonds as nodes, where time evolution follows the (inverse) tensor network renormalization group of Evenbly and Vidal~\cite{EvenblyVidal2015}. Because this holographic space-time approach relies fundamentally on finite-dimensional subsystems from the outset, the typicality bounds of Lubkin and Page directly apply, bypassing the need for infinite-dimensional algebraic generalizations.

While such finite-dimensional frameworks elegantly circumvent the issue by explicitly converting continuous Type~III algebras to discrete Type~I$_N$ systems, whether a fully algebraic Lubkin-Page theorem can be formulated natively for Type~III factors in the continuum limit is nevertheless one of the central open problems at the
interface of algebraic quantum field theory and quantum gravity.
If the Bekenstein-Hawking bound renders the physical Hilbert space
finite-dimensional, however, or as explicitly realized in the hydrodynamic approach discussed above, the Type~III structure is an artefact
of the continuum approximation, and the results proved here and
in~\cite{WangBraunstein2025PRL} apply directly.

\end{document}